\documentclass[10pt]{article}

\usepackage{a4wide}
\usepackage{amssymb}

\begin{document}

\def\O{{\cal O}}
\def\N{{\cal N}}
\def\>t{>_{\scriptscriptstyle{\rm T}}}
\def\enu{\epsilon_\nu}
\def\pint{\int{\d^3p\over(2\pi)^3}}
\def\gint{\int[\D g]\P[g]}
\def\nbar{\overline n}
\def\d{{\rm d}}
\def\e{{\bf e}}
\def\x{{\bf x}}
\def\y{{\bf y}}
\def\hn{{{\bf h}_n}}
\def\hm{{{\bf h}'_m}}
\def\hp{{{\bf h}_p}}
\def\0x{\x^\smalze}
\def\sperpx{{x_\perp}}
\def\sperpk{{k_\perp}}
\def\sbperpk{{{\bf k}_\perp}}
\def\sbperpx{{{\bf x}_\perp}}
\def\perpx{{x_{\rm S}}}
\def\perpk{{k_{\rm S}}}
\def\bperpk{{{\bf k}_{\rm S}}}
\def\bperpx{{{\bf x}_{\rm S}}}
\def\p{{\bf p}}
\def\q{{\bf q}}
\def\zr{{\bf z}}
\def\R{{\bf R}}
\def\A{{\bf A}}
\def\v{{\bf v}}
\def\u{{\bf u}}
\def\w{{\bf w}}
\def\U{{\bf U}}
\def\cm{{\rm cm}}
\def\l{{\bf l}}
\def\sec{{\rm sec}}
\def\Ckol{C_{Kol}}
\def\flux{\bar\epsilon}
\def\zq{{\zeta_q}}
\def\b{b_{kpq}}
\def\bun{b^{\scriptscriptstyle (1)}_{kpq}}
\def\bdu{b^{\scriptscriptstyle (2)}_{kpq}}
\def\z0q{{\zeta^{\scriptscriptstyle{0}}_q}}
\def\smalS{{\scriptscriptstyle {\rm S}}}
\def\smalze{{\scriptscriptstyle (0)}}
\def\smalI{{\scriptscriptstyle {\rm I}}}
\def\smalun{{\scriptscriptstyle (1)}}
\def\smaldu{{\scriptscriptstyle (2)}}
\def\smaltr{{\scriptscriptstyle (3)}}
\def\smalL{{\scriptscriptstyle{\rm L}}}
\def\smalD{{\scriptscriptstyle{\rm D}}}
\def\smal1n{{\scriptscriptstyle (1,n)}}
\def\smaln{{\scriptscriptstyle (n)}}
\def\smalA{{\scriptscriptstyle {\rm A}}}
\def\shell{{\tt S}}
\def\ball{{\tt B}}
\def\nav{\bar N}
\def\micron{\mu{\rm m}}
\font\brm=cmr10 at 24truept
\font\bfm=cmbx10 at 15truept
\centerline{\brm Intermittency, synthetic turbulence}
\centerline{\brm and wavelet structure functions}
\vskip 20pt
\centerline{Piero Olla$^1$ and Paolo Paradisi$^2$}
\vskip 5pt
\centerline{$^1$ISIAtA-CNR, 73100 Lecce Italy}
\centerline{$^2$FISBAT-CNR, 40129 Bologna Italy}
\vskip 20pt

\centerline{\bf Abstract}
\vskip 5pt
Some techniques for the study of intermittency by means of wavelet transforms, are 
presented on an example of synthetic turbulent signal. Several features of the turbulent 
field, that cannot be probed looking at standard structure function scaling, become 
accessible in this way. The concept of a directly measurable intermittency 
scale, distinct from the scale of the fluctuations, is introduced. A method for optimizing 
the analyzing wavelets, which exploits this concept, and allows to minimize non-local 
contributions in scale to wavelet correlations, is described. The transition from a 
wavelet to a Fourier transform based description of an intermittent random field, and the
possibility of using Fourier correlations to measure intermittency are discussed. 
Important limitations in the ability of structure functions to give a local in scale 
description of intermittent random fields, are observed.

\vskip 2cm
\centerline{\it Submitted to}
\centerline{\it Physica D}
\centerline{\it 3/27/98}
\vfill\eject
\centerline{\bf I. Introduction}
\vskip 5pt
Structure functions and scaling arguments, are basic tools in the study of turbulence.
One of the key issues in the field, intermittency, is expressed in this language through
the statement, that the scaling of the structure functions $S_l(q)=\langle|\Delta_l v|^q\rangle$,
with $\Delta_lv(x)=v(x+l)-v(x)$, is non-trivial. This means simply, that in high Reynolds
number turbulence, there is a large ''inertial'' range of scales where $S_l(q)$ obeys a 
power law: $S_q(l)\sim l^{\zeta_q}$, with $\zeta_q$ a nonlinear function of the order $q$. 

Many explanations have been proposed for this phenomenon, but to date, nothing definitive
is still available (see \cite{frisch} for a review). There are several reasons for this state 
of affairs; one is 
perhaps, that scaling by itself is not able to provide a sufficiently complete description,
of what is happening in the turbulent field. Within Kolmogorov 1941 theory \cite{K41}, 
energy
conservation is able by itself to fix a value of the scaling exponent: $\zeta_q^\smalze=q/3$,
which is an acceptable lowest order approximation for $\zeta_q$. However, no comparable symmetry
based argument has been proposed, which was able to fix the value of the intermittency
correction $\zeta_q^\smalze-\zeta_q$.

The fact is that, although the mechanism of intermittency generation is 
likely to be universal, possibly associated with some property of the nonlinear energy
transfer in the inertial range, it is also likely that its effect is not exhausted 
in the production of scaling corrections. In fact, very different kinds of intermittency,
like e.g. the one associated with the long and thin vortices observed in numerical 
simulations \cite{vincent91}, and the one that would be obtained in a random beta-model 
picture of turbulence \cite{benzi85}, produce equally acceptable spectra of scaling 
corrections.

Wavelet analysis \cite{daubechies92} has often been proposed as an alternative tool in the 
study of turbulent
intermittency. The possibility of having an additional degree of freedom associated with
the wavelet shape, beyond position in space and scale, makes these objects, particularly
appropriate to study coherent structures and the geometrical properties of intermittency
\cite{farge92}.
However, the great freedom to describe geometrical features in two- or three-dimensional
settings, thus becoming available, has hindered perhaps an extensive use of these 
techniques in turbulence theory.

An alternative use of wavelets has been to employ them as building blocks in the generation
of artificial turbulent signals, trying to reproduce the kind of velocity time series
one gets in experiments. This kind of technique, first introduced in \cite{eggers91} 
to study connections between dissipation and velocity intermittency, and in \cite{vicsek91},
has been later developed 
systematically in \cite{benzi93} (see \cite{juneja94} for recent references). More recently 
\cite{olla97}, this approach has been used to provide a kinematic explanation for the 
kind of energy spectra developing in wall turbulence. 
The picture of a superposition of eddies at different scales obtained
in this way, appears particularly natural to study intermittency effects. However, even
in such one-dimensional settings, there is a great freedom in the 
choice of the eddy generation mechanism. It is clear, for example, that the same spectrum 
of anomalous exponents can be obtained, both from a ''random eddy model'' with inclusion 
of intermittency, like the one considered in \cite{olla97}, and 
from a multiplicative cascade of the kind described in \cite{benzi93}. (To get such an 
identically scaling signal, out of the first model, it is enough to randomly permute the  
wavelets at the different scales in the second).
Now, different mechanisms of turbulence synthesis imply, to some degree, different 
assumptions on the real turbulent dynamics. It is therefore of some relevance, to devise
methods which allow to identify these mechanisms, from the statistical properties of the
signal.

In light of the present discussion, synthetic turbulence appears to be the appropriate 
''training facility'', in which to test different techniques of wavelet analysis, and
their ability to reveal specific intermittency features \cite{arneodo97}. 

Recently, spatial \cite{arneodo98} and scale \cite{benzi98} correlations between wavelet 
components, have been used to probe the cascade structure of turbulent signals. Here,
the interest is focused on two different issues: the choice of the wavelet in both the
signal generation and analysis, and the amount of phase space available to each ''building
block'' wavelet in the generation algorithm. This phase space is given by the relative 
position and scale of the generated wavelet, with respect to the parent one, and in 
typical algorithms of signal generation \cite{benzi93}, it consists just of a single point.
It turns out that both issues of wavelet shape and phase space availability, have important
consequences as regards the ability of a structure function to detect features of the
turbulent dynamics that are really local in scale.

In the next section, the generation algorithm for the synthetic turbulent signal is introduced,
and in section III, the basic wavelet structure function properties are derived. A systematic
analysis of the turbulent statistics dependence, on the shape of both
the analyzing and the building block wavelets, is carried on in section IV.
Given the one-dimensional nature of the signal,
geometrical aspects are minimal, and this leaves out only one essential degree of freedom
in the choice of the wavelets: their ''number of wiggles'', i.e. the product 
of their dominant wavevector and spatial extension. 
Section V contains discussion of the results and conclusions. We leave in the appendix,
the analysis of the case in which the cascade is probabilistic in space and discrete in scale.
\vskip 20pt

\centerline{\bf II. Synthesis of an artificial turbulent signal}
\vskip 5pt
Introduce the Gaussian wavepacket:
$$
w_S(k,y,x)=\exp(ik(x-y)-(x-y)^2/\lambda^2_S);\qquad\lambda_S=a_S/k
\eqno(1)
$$
and let:
$$
\Psi(x)=\sum_n\sum_\hn A_\hn w_S(k_\hn,y_\hn,x)
\eqno(2)
$$
be a real random field, which should mimic the time signal from a fixed position
velocity measurement in a turbulent flow. Following standard practice \cite{benzi93}, 
the building block wavelets $w_S$ are generated through a cascade process,
to model the mechanism of energy transfer in the turbulent flow.
The vector index $\hn=(h_0,h_1,....h_n)$ identifies then the position of the wavelet
in the cascade through the sequence of its ancestors: the integer $h_n$ labels the 
$h_n$-th daughter wavelet generated at the $n$-th step in the cascade, by wavelet 
${\bf h}_{n-1}$. 

Notice, however, that an intermittent random field could be generated, without any 
reference to cascade processes, either 
by varying appropriately the space density of the wavelets with scale, or by 
making the distribution of the amplitudes $A$ more intermittent as $k$ grows, 
but keeping the wavelets randomly distributed in space \cite{olla97}.

From reality of $\Psi$, for each index $\hn$ with positive 
components, there is a wavelet with index $-\hn$ such that $A_{-\hn}=A_\hn^*$ 
$y_\hn=y_{-\hn}$ and $k_\hn=-k_{-\hn}$. The cascade is assumed
to be local in $n$ in the sense that the probability that a given wavelet 
$\hn$ has a certain value of its parameters $\xi_\hn\equiv\{\ln A_\hn,\ln k_\hn,y_\hn\}$,
can be written in terms of transition probabilities, as: 
$$
P(\xi_{h_0}\to\xi_\hn)=
\int\d\xi_{{\bf h}_1}...\d\xi_{{\bf h}_{n-1}} p(\xi_{h_0}\to\xi_{{\bf h}_1})...
p(\xi_{{\bf h}_{n-1}}\to\xi_\hn).
\eqno(3)
$$
For the sake of simplicity, the transition probabilities are assumed to factorize
into their $\ln A$, $\ln k$ and $y$ components, with scale invariance forcing the
cascade to be governed by a multiplicative random process:
$$
p(\xi\to\xi')=p_A(A'/A|k'/k)p_k(k'/k)p_x(k|y-y'|).
\eqno(4)
$$
The relative phase of $A'$ and $A$ is assumed random, and we take:
$$
\int\d\ln y\, p_A(y|x)y^p=c_px^{-\zeta_p},
\eqno(5)
$$
in order to get power law scaling in the structure functions. Given Eqns. (4-5), the transition 
probability over $n$ steps is in the form:
$$
P(\xi_\hn\to\xi_\hp)=P_A(p-n,A_\hp/A_\hn|k_\hp/k_\hn)P_k(p-n,k_\hp/k_\hn)P_x(k_\hn|y_\hp-y_\hn|).
\eqno(6)
$$
At each step $n$ in the cascade, the wavelets distribution in scale is peaked at 
$\bar k_n=\bar k_0\exp(n\bar z)$, with the wavevectors $\bar k_0$ distributed around a 
characteristic large scale $L$: $\langle k_0\rangle=L^{-1}$. Each mother wavelet generates 
$\exp(\bar z)$ daughters; this insures that the mean degree of overlap between wavelets in 
$k$-$y$ space is scale invariant. 

There are several reasons to consider a mechanism of turbulence synthesis, in which the 
wavelets are distributed in space and scale in a probabilistic way, rather than 
on a rigid lattice. The main, rather ''philosophical'' motivation, however, is to try 
considering the building block wavelets, more like eddies (or components of bunches of 
eddies if $a_S$ is large), than like basis functions at fixed position in space; this 
also in view of possible extensions of the model to time dependent situations, in which 
the eddies are mobile. 
\vskip 20pt

\centerline{\bf III. Analysis of an artificial turbulent signal}
\vskip 5pt
The choice of analyzing wavelet is in general arbitrary. A Gaussian wavepackets, however,
has the minimum spread in $k-y$ space and allows to retain the maximum simultaneous 
information possible in space and scale. We thus take from the start the analyzing wavelets
to be derivatives of Gaussian wavepackets: $(-k)^{-s}\partial_x^sw_A(k,y,x)$, for which in 
general: $a_A=\lambda k\ne a_S$.
The components of $\Psi(x)$ with respect to this set of wavepackets are defined as follows:
$$
\Psi_{ky}=\lambda_A^{-1}k^{-s}\int\d x\, w_A^*(k,y,x)\partial_x^s\Psi(x).
\eqno(7)
$$
together with the associated structure functions $\langle|\Psi_{ky}|^q\rangle$.
In order for these structure functions not to be dominated by the largest scales in $\Psi(x)$,
given standard Kolmogorov scaling for $\Psi$, it is necessary that the parameter $s$ in Eqn. 
(7) be at least equal to one. To calculate them , we need to evaluate first the component
of a building block wavelet on an analyzing wavelet. Because of random phase of $A$, we will
need only its square modulus: $C(k,y;k',y')=|k^{-m}\int\d x\, 
w_A^*(k,y,x)\partial_x^mw_S^*(k',y',x)|^2$. If the probabilities $P_k$ vary sufficiently slow, 
and $s$ is large enough to kill the contribution to the $k$-integrals from $k$ small,
it is possible to write from the start:
$$
C(k_A,y_A;k_S,y_S)={a_S^2\over a_A^2+a_S^2}
\exp\Big(-{2 k_A^2\over a_A^2+a_S^2}\Big(\Delta y^2+{a_Aa_S\Delta k^2 \over 4k_A^2}\Big)\Big)
\eqno(8)
$$
with $\Delta k=k_A-k_S$ and $\Delta y=y_A-y_S$, and the integrals over $k$, that arise in 
the averages involved in the correlations can be carried out by steepest descent, disregarding 
the contribution at $k\sim L^{-1}$. 

The simplest correlations are $\langle |\Psi_{ky}|^2\rangle$ and  
$\langle |\Psi_{ky}|^2|\Psi_{k'y'}|^2\rangle$. 
Given the random phase of $A$, it is easy to see from Eqns. (2) and (8), that a $2n$-order 
correlation will receive contribution at most
by $n$ eddies, and the $4$-th order correlation will be in the form:
$$
\langle |\Psi_{ky}|^2|\Psi_{k'y'}|^2\rangle
=\langle |\Psi_{ky}|^2|\Psi_{k'y'}|^2\rangle_1+\langle |\Psi_{ky}|^2|\Psi_{k'y'}|^2\rangle_2
$$
$$
=\sum_nF_n\langle |A_\hn|^4\rangle+\sum_{np}G_{np}\langle |A_\hn|^2|A_\hp|^2\rangle.
\eqno(9)
$$
From the form of $p_A$, we have power laws for the amplitude correlations;
$$
\langle |A_\hn|^2\rangle=c_2(k_\hn L)^{-\zeta_2}
\eqno(10)
$$
and
$$
\langle |A_\hn|^2|A_\hm|^2\rangle=c_4(k_\hm L)^{-\zeta_4}(k_\hn/k_\hm)^{-\zeta_2}
(k_\hm/k_\hp)^{\zeta_4-2\zeta_2}
\eqno(11)
$$
with $k_\hn>k_\hm$ and $p$ the cascade step at which the genealogical tree of $\hn$ and $\hm$ 
branches: $h_i=h'_i$ for $0\le i\le p$ and $h_i\ne h'_i$ for $i>p$. Thus, the lower the
branching takes place in the tree, the closer the correlation gets to its disconnected
limit $\langle |A_\hn|^2|A_\hm|^2\rangle=c_4 (k_\hm k_\hn L^2)^{-\zeta_2}$.

From Eqns. (2) and (7) we obtain, for the second order correlation:
$$
\langle |\Psi_{ky}|^2\rangle=
c_2\int\d\bar y\d\ln \bar k\sum_n\bar k_n\langle P_k(n,\bar k/k_0)\rangle (\bar kL)^{-\zeta_2}
C(k,y;\bar k,\bar y),
\eqno(12)
$$
where $\langle P_k(n,\bar k/k_0)\rangle$ is averaged over $k_0$ (from summing over $h_0$ in $\hn$),
and $C(k,y;k',y')$ the square modulus of the component of wavelet $w_S(k',y',x)$ with respect to the
analyzing wavelet $w_A(k,y,x)$:
$C(k,y;k',y')=|k^{-m}\int\d x\, w_A^*(k,y,x)\partial_x^mw_S^*(k',y',x)|^2$. 

We have an analogous expression for the one-eddy contribution
to $\langle |\Psi_{ky}|^2|\Psi_{k'y'}|^2\rangle$:
$$
\langle |\Psi_{ky}|^2|\Psi_{k'y'}|^2\rangle_1
$$
$$
=c_4\int\d\bar y\d\ln\bar k\sum_n\bar k_n\langle P_k(n,\bar k/\bar k_0)\rangle (\bar kL)^{-\zeta_4}
C(k,y;\bar k,\bar y)C(k',y';\bar k,\bar y).
\eqno(13)
$$
For the two-eddy contribution, we have instead, for $k'\ge k$:
$$
\langle |\Psi_{ky}|^2|\Psi_{k'y'}|^2\rangle_2
=2c_4\int\d\ln\bar k\d\ln\bar k'\d\bar y\d\bar y'\int_{\ln L^{-1}}^{\ln k}\d\ln\hat k\,
(\bar kL)^{-\zeta_4}(\bar k'/\bar k)^{-\zeta_2}
(\bar k/\hat k)^{\zeta_4-2\zeta_2}
$$
$$
\times\sum_n\sum_{m=0}^n\sum_{p=0}^m{\bar k_n\bar k_m\over\bar k_p}\langle P_k(p,\hat k/\bar k_0)
\rangle P_k(n-p,\bar k/\hat k)P_k(m-p,\bar k'/\hat k)
$$
$$
\times P_x(\hat k|\bar y-\bar y'|)C(k,y;\bar k,\bar y)C(k',y';\bar k',\bar y').
\eqno(14)
$$
where $(\bar k_m/\bar k_p)P_x(\hat k|\bar y-\bar y'|)$ gives the space
density at $\bar y'$ of eddies $\hm$ generated from the branching at $\hp$, given the
presence of an eddy $\hn$ at $\bar y$. The factor $\bar k_m/\bar k_p=\exp((m-p)\bar z)$ is the
actual number of eddies $\hm$ generated from the branching at $\hp$. The factors $\bar k_n$
entering Eqns. (12-14), conversely, give the space density of wavelets of typical size 
$a_S/\bar k_n$, at the $n$-th step in the cascade.

The cascade structure is characterized by a discrete component through the sums entering 
Eqns. (12-14). Since the cascade steps are independent, the width of the cumulative 
distribution $P_k(n,k'/k)$ is $n^{1\over 2}\Delta z$ with $\Delta z$ the width of 
$p_k$: $\Delta z^2=\int\d\ln k'|\ln k'/k-\bar z|^2p_k(k'/k)$. Thus, if the separation of the 
scales entering the $P_k$ involved in the sums in Eqns. (12-14) is large enough, the
effect of discreteness will be negligible. The same will occur if $\Delta z/\bar z$ 
itself, is large enough. In the other limit, when $\Delta z/\bar z$ and $n$ 
are small, oscillation with period $\bar z$ in $\ln k'/k$ and $\ln kL$ (lacunarity)
are to be expected in the scale dependence of the correlations.
\vskip 20pt

\centerline{\bf IV. Structure functions and optimization of the analyzing wavelets}
\vskip 5pt
The length $\lambda_S$ introduced in Eqn. (1) plays an important role in the generation 
of an intermittent random field, since it identifies the intermittency scale of 
fluctuations of size $k^{-1}$. If $a_S$ is large, indeed, it is not the single fluctuation
that is intermittent, but the amplitude of a whole bunch of them, extended over a length
$\lambda_S$. As a consequence of this, one expects that, if $\lambda_S$ approaches the size
of the domain, intermittency is lost and a standard, random phase generated gaussian
field is obtained. Conversely, if one carries on the same operation with the length 
$\lambda_A$, the final result is that, instead of dealing with structure functions, 
one ends up working with correlations between Fourier components of the random field.
Again, intermittency is expected to be lost. If one is interested in studying 
intermittency by structure function scaling, the optimal choice should be therefore: 
$\lambda_A\sim\lambda_S$. For this reason, it becomes necessary to study the dependence 
of the structure functions $\langle |\Psi_{ky}|^p\rangle$ on the parameters $a_A$ and $a_S$. 

If $\Delta z/\bar z$ is not too small, discreteness effects in scale can be neglected and
the sums in Eqns. (12-14) can be approximated by integrals.
The transition probability $P_k$ is in the form: $P_k(n,k'/k)=f(n,\ln k'/k-n\bar z)\simeq
f(\bar z^{-1}\ln k'/k,\ln k'/k-n\bar z)$. Hence:
$$
\sum_{n=1}^\infty (\bar k_n/k)P_k(n,k/\bar k_0)\simeq\int\d x\bar z^{-1}
\e^xf(\bar z^{-1}\ln k/k_0,x)= {1\over\bar z}\Big({k\over\bar k_0}\Big)^\epsilon,
\eqno(15)
$$
where $\epsilon=\lim_{n\to\infty}{1\over n\bar z}\ln\langle\exp(x)|n\rangle=
\O(\bar z^{-3}\Delta z^2)$
and $\langle...|n\rangle$ indicates average over $f(n,x)$. We thus see that scale uncertainty in 
the process of eddy generation contributes to scaling in such a way that $\zeta_q\to\zeta'_q=
\zeta_q+\epsilon$.

At this point, all the terms in $P_k$ and associated sums drop off Eqns. (12-14) and it is
possible to evaluate the integrals for $k=k'$ and $y=y'$. From Eqns. (12-13), we get immediately:
$$
\langle |\Psi_{ky}|^2\rangle\simeq{c_2\pi a_S\over \bar za_A}(kL)^{-\zeta'_2}
\quad{\rm and}\quad
\langle |\Psi_{ky}|^4\rangle_1\simeq{c_4\pi a_S^3(kL)^{-\zeta'_4}\over \bar za_A(a_S^2+a_A^2)}
\eqno(16)
$$
Performing some power counting
on Eqn. (14), however, we discover that the $\d\hat k$ integral is dominated more and more
by large scales the closer we are to trivial scaling: $\zeta'_4=2\zeta'_2$. In this regime,
if $p_x$ is sufficiently well behaved, it will be possible to approximate $P_x$, from
central limit theorem arguments, with a Gaussian: 
$$
P_x(k|x-x'|)\simeq{k\over\pi^{1\over 2}\hat ba_S}\exp\Big(-{k^2|x-x'|^2\over \hat b^2a_S^2}\Big)
\eqno(17)
$$
with the parameter $\hat b$ characterizing the spatial non-locality of the cascade. Substituting 
into Eqn. (14), we obtain:
$$
\langle |\Psi_{ky}|^4\rangle=\langle |\Psi_{ky}|^4\rangle_1+\langle |\Psi_{ky}|^4\rangle_2
$$
$$
\simeq{c_4\pi a_S^2\over \bar za_A^2}(kL)^{-\zeta'_4}
\left[ {a_Sa_A\over a_S^2+a_A^2}+{2\pi^{1\over 2}\over \bar z^2\hat ba_S}
\int_0^{\ln kL}\d x\Big(1+{a_A^2+a_S^2\over\hat b^2a_S^2}\e^{-2x}\Big)^{-{1\over 2}}
\e^{(\zeta'_4-2\zeta'_2)x}\right]
\eqno(18)
$$
This equation is our main result, and tells us how the various eddies in the synthetic turbulent 
field, contribute to structure function scaling.

The second term in square brakets in Eqn. (18) is the two-eddy part of
$\langle |\Psi_{ky}|^4\rangle$, giving for each $x=\ln k/\hat k$, the logarithmic distance of 
the common ancestor of the two eddies from the scale $k$. For $\zeta'_4-2\zeta'_2$ small, which 
is true in the case of turbulence, this integral receives contributions from 
$\max(0,{1\over 2}\ln{a_A^2+a_S^2\over\hat b^2a_S^2})<x<\ln kL$, where the integrand can be
approximated by $\exp((\zeta'_4-2\zeta'_2)x)$. We have then the important 
result, that for $kL<\exp((2\zeta'_2-\zeta'_4)^{-1})$, which is a very large range
of scales, the structure function contains a logarithmic two-eddy contribution, which comes 
right from the largest scales in the random field. We obtain then, for the kurtosis
$K_4(k,a_S,a_A)=\langle |\Psi_{ky}|^2\rangle^{-2}\langle |\Psi_{ky}|^4\rangle$:
$$
K_4(k,a_S,a_A)\simeq {c_4\bar z\over\pi c_2^2}(kL)^{2\zeta'_2-\zeta'_4}
\left[{a_Sa_A\over a_S^2+a_A^2}+{2\pi^{1\over 2}R^{2\zeta'_2-\zeta'_4}\over \bar z^2\hat ba_S}
\ln RkL\right]
\eqno(19)
$$
where: $ R=\max(1,({\hat b^2a_S^2\over a_A^2+a_S^2})^{1\over 2})$.
Only for $kL\gg\exp((2\zeta'_2-\zeta'_4)^{-1})$ we reach pure power law scaling:
$$
K_4(k,a_S,a_A)\simeq {c_4\bar z\over\pi c_2^2}(kL)^{2\zeta'_2-\zeta'_4}
\left[ {a_Sa_A\over a_S^2+a_A^2}+{2\pi^{1\over 2}R^{2\zeta'_2-\zeta'_4}\over 
\bar z^2\hat ba_S(2\zeta'_2-\zeta'_4)}\right]
\eqno(20)
$$
If, instead of looking at the scaling of $K_4(k,a_S,a_A)$, we study the
dependence of this quantity on $a_A$ for $k$ fixed, we notice the presence of a maximum
at $a_A=a_S$. This corresponds to the maximum possible overlap between building block 
and analyzing wavelets; for $a_A>a_S$, an analyizing wavelet will feel the effect of many
eddies at different position in space, while, for $a_A<a_S$, these will be distributed at 
different scales. It is important to notice, as it is clear from Eqn. (19), that this
effect will be felt also in the measured scaling exponents, that, because of the logarithm,
will be dependent 
on $a_A$. This effect will be minimum only at $a_A=a_S$, when the local in scale, 
one-eddy contribution to the structure function is maximum. Conversely, the importance
of the two-eddy contribution goes to zero when $a_S$ is large.

It is important to stress the importance of the smallness of $2\zeta'_2-\zeta'_4$ and of the
cascade structure of the random field. This causes the slow decay of the two-eddy contribution 
as $a_A$ gets large. In a random eddy model of the kind considered in \cite{olla97}, the
two-eddy contribution would scale like $(kL)^{-2\zeta'_2}$ whatever the value of $a_A$,
the reason being the lack of correlations among eddies. 

When $a_A/a_S$ becomes large enough, the one-eddy contribution can be disregarded, and the
kurtosis $K_4$ begins to scale in $R$, i.e. in the ratio $\lambda_S/\lambda_A$, so that we have,
to leading order in $k$:
$$
K_4(k,a_S,a_A)\sim (L/\lambda_A)^{2\zeta'_2-\zeta'_4}
\eqno(21)
$$
In fact, for $a_A$ large, $w_A$ probes eddies coming from common ancestors, which can be very 
distant in scale from $k$, and which become uncorrelated when $\lambda_A>L$.
From Eqn. (21), $\langle |\Psi_{ky}|^4\rangle$ obeys trivial scaling:
$\langle |\Psi_{ky}|^4\rangle\sim (kL)^{-2\zeta'_2}$ and the intermittent nature of the
random field is lost.

In this regime, the ratio of the one- to two-eddy contribution to $\langle |\Psi_{ky}|^4\rangle$
scales, again to leading order in $k$:
$$
{\langle |\Psi_{ky}|^4\rangle_1\over\langle |\Psi_{ky}|^4\rangle_2}\sim 
(kL)^{-1-\zeta'_4+2\zeta'_2}
\eqno(22)
$$ 
The factor $(kL)^{-1}$ in this equation has a very important interpretation when we consider
that, for fixed $\lambda_A\sim L$, and $kL$ large, $\lambda_A\Psi_{ky}\to\Psi_k$, which is the 
Fourier transform of $\Psi(x)$ in a box of size $L$. We have then, that Eqns. (21-22)
take the form: $\langle |\Psi_k|^4\rangle\sim aLk^{-3}(Lk)^{-\zeta'_4}+bL^2k^{-2}(Lk)^{-2\zeta_2}$
which is nothing else than the expression $\langle \Psi_{k_1}\Psi_{k_2}\Psi_{k_3}\Psi_{k_4}\rangle
=\delta(k_1+k_2+k_3+k_4)C_4(k_1,k_2,k_3,k_4)+\delta(k_1+k_2)\delta(k_3+k_4)C_2(k_1)C_2(k_3)$
for $k_1=-k_2=k_3=-k_4$ with $\delta(0)\sim L$. Thus, $\langle |\Psi_{ky}|^4\rangle_2$
kill $\langle |\Psi_{ky}|^4\rangle_1$ by a factor $kL$ that is the term $k\delta(0)$ that gives
the ratio of the disconnected to the connected Fourier 4-point correlation of a uniform random 
field. All this suggests that, perhaps, structure functions are not the most appropriate objects 
to probe features of the field which are local in scale; rather, the multiscale Fourier 
correlation: $\langle\Psi_{k_1}\Psi_{k_2}\Psi_{k_3}\Psi_{k_4}\rangle$ with $k_i\ne -k_j$
$\forall\ i,j$ and $k_1+...+k_4=0$ should be taken into consideration:
$$
\langle \Psi_{k_1}...\Psi_{k_4}\rangle
\simeq 2\pi\delta(k_1+...+k_4){\pi^2a_S^4\over \bar z}
\int{\d k\over k^4}(kL)^{-\zeta'_4}\exp\Big(-{a_S^2\over 4k^2}\sum_{n=1}^4(k_n-k)^2\Big).
\eqno(23)
$$
Only in this way, it would be possible to avoid logarithmic corrections
in the scaling of correlation functions, coming from two-eddy effects.
\vskip 20pt

\centerline{\bf V. Conclusions}
\vskip 5pt
The motivation for the interest in synthetic turbulence has often been, more in the
''output'', i.e. in the turbulent field or turbulent signal being produced, than in
the dynamical meaning of the adopted algorithm. A typical application has been,
for instance, the possibility of controlling velocity spectra, in the study of
turbulent diffusion \cite{fung92}. When it comes to an issue like intermittency, however,
the interest is more in the algorithm itself and in the effect that different choices for
it, would have on the turbulent statistics. In the present research, the main result
is the ease, with which some innocent looking algorithms
for the generation of synthetic turbulence, lead to strongly non-local effects in structure
function scaling. This, despite the local nature of the cascade mechanism on which the algorithms
are based.

The only way this phenomenon can be
explained is through the interaction of the cascade nature of the 
algorithm, with the probabilistic distribution in space and scale of the eddies.
In particular, had the eddies been distributed at random, without a cascade structure, 
the two eddy contribution to a 4-th order structure function, would have been trivially:
$\int\d\bar k\d\bar k'\d\bar y\d\bar y'C(k,y;\bar k,\bar y)C(k,y;\bar k',\bar y')$
$\times(\bar k\bar k')^{-\zeta_4/2}\sim k^{-\zeta_4}$, with $C(k,y;\bar k,\bar y)$ the square
wavelet component of an eddy $[$see Eqn. (9)$]$.
On the other hand, if the cascade had been rigid, organized on a lattice structure as in 
\cite{benzi93}, this contribution would have been simply $NC(k,y;\bar k,\bar y)^2
\bar k^{-\zeta_4}\sim k^{-\zeta_4}$, with $N$ the number of wavelets $w_S$ in 
the lattice, overlapping with the wavelet $w_A$, and $\bar k\bar y$ the typical scale and
coordinate of the wavelets $w_S$. 

All these effects manifest themselves in the behavior of structure functions, through 
logarithmic scaling corrections carrying information on the largest scales of the signal.
One result of the present study could therefore be that, if one wanted simply to generate a 
random field with prescribed multifractal statistics, it would not be a good idea to include 
probabilistic effects in the space and scale distribution of eddies.  If, on the other hand,
one thinks that all this may have some relevance for real turbulence, the conclusion is 
that there exists a kinematic effect, which could limit the ability of objects like
structure functions, to detect scale-local features of the turbulent field. This 
confirms the current thinking on the subject, which leads to expect non-locality, among
the other things, from the convolution nature, in Fourier space, of correlations of order
greater than two.  It is worth stressing the kinematic nature of our result, which is due 
to the way in which contributions from different eddies sum up in structure functions: 
the dynamics of the eddy generation mechanism remains strictly local in scale.

The question at this point is how to verify that the non-locality effects we have discussed
are of any relevance in high Reynolds numbers turbulence. An experimental test could be 
the dependence of the wavelet structure function scaling exponents, on the parameter $a_A$: 
the number of wiggles of the wavelet. In this case, the suggestion from the present research, 
is that there should be a 
value of $a_A$, for which non-local effects become minimal, and which is identified
by the maximum at fixed scale of the generalized kurtosis $[$see Eqns. (19-20)$]$.
This maximum has a nice physical interpretation in terms of resonance between 
the analyizing wavelet and intermittency, with the wavelet dominant wavevector 
identifying the scale of the turbulent 
fluctuations, and the wavelet extension giving the lengthscale over which these fluctuations,
act coherently to generate intermittency. If the picture of a probabilistic cascade is right,
however, only a description based on Fourier correlation scaling, could allow elimination of 
this kind of non-local effects, once all disconnected contributions to the correlation, are 
eliminated by appropriate choice of the wavevectors.

The results of this study are of rather general validity, showing that there is a 
rather broad class of intermittent random fields, for which techniques based on the 
analysis of structure function scaling, are of limited use.
If the interest is more in the modelling of turbulence intermittency, however, our 
non-locality effects may be considered more as an artifact of the algorithm of turbulence 
generation. On the other hand, it is difficult to a priori exclude a probabilistic 
cascade, as opposed to a ''rigid'' one, and several arguments in favor and against both are
easy to find. A drawback of a probabilistic cascade is that the eddies can overlap 
in $k-y$ space, while being treated as independent objects in the random multiplicative process. 
One may argue back, however, that we are dealing with a one dimensional section of a 
three-dimensional turbulent field, and that these overlaps are therefore irrelevant. Conversely, 
it is more aesthetically pleasing to distribute the wavelets in the random field, freely in 
$k-y$ space, but then one looses the property of the wavelets, of being base functions
for the random field. 

In any case, there are practical reasons for being interested in probabilistic cascades.
One is the possibility of studying the effect of space and scale non-locality in the eddy 
generation \cite{olla98}, which have a direct interpretation in terms of properties of the 
energy transfer in real turbulence. The second is, that the lack of constrains over the 
eddy position allows, in a time dependent situation, to put these ''eddies'' in motion, 
accounting for sweep in time correlations, in a much easier way than using fixed wavelets.

\vskip 10pt
\noindent{\bf Aknowledgements}: I would like to thank Jean-Fran\c cois Pinton, Sergio
Ciliberto and Jens Eggers for interesting and stimulating discussion.
\vskip 20pt

\centerline{\bf Appendix: The effect of discreteness in the cascade}
\vskip 5pt
For the sake of completeness, we examine the discrete limit of
Eqn. (14), analyzing how, restriction of the phase space available to the eddies at their birth,
modifies the scale non-local character of the struture functions.

An argument analogous to the one used to arrive at Eqn. (17), leads to the expression for
$P_k$:
$$
P_k(n;k'/k)={1\over (n\pi)^{1\over 2}\Delta z}
\exp\Big(-{(\ln(k'/k)-n\bar z)^2\over n\Delta z^2}\Big).
\eqno(A1)
$$
The continuous approximation, used to arrive from Eqns. (12-14) at Eqns. (16) and (18),
can still be applied to the sums in $n$ of Eqns. (12-13) and in $p$ of 
Eqn. (14), also for $\Delta z/\bar z$ small. This, provided  $\Delta z^2\ln kL/\bar z>1$. The 
effect of discreteness remains therefore in the sums over $n$ and $m$ in Eqn. (14), which, for
$\Delta z/\bar z$ small, are dominated by $n=p+{\rm Int}(\bar z^{-1}\ln\bar k/\hat k)$ and
$m=p+{\rm Int}(\bar z^{-1}\ln\bar k'/\hat k)$. From Eqn. (14), we thus obtain the following 
limit expression for $\langle |\Psi_{ky}|^4\rangle_2$:
$$
\langle |\Psi_{ky}|^4\rangle\simeq
{2c_4\over\pi\Delta z^2}\int\d\bar k\d\bar k'\d\bar y\d\bar y'
\int_{L^{-1}}^{\bar k}{\d\hat k\over \hat k^2}
{(\bar kL)^{-\zeta'_4}(\bar k'/\bar k)^{-\zeta'_2}(\bar k/\hat k)^{\zeta'_4-2\zeta'_2}
\over (\ln(\bar k/\hat k)\ln(\bar k'/\hat k))^{1\over 2}}
$$
$$
\times P_x(\hat k|\bar y-\bar y'|)C(k,y;\bar k,\bar y)C(k,y;\bar k',\bar y')
\exp\Big(-{\bar z^3\over\Delta z^2}\Big({\Delta_{\hat k\bar k}^2\over\ln\bar k/\hat k}+
{\Delta_{\hat k\bar k'}^2\over\ln\bar k'/\hat k}\Big)\Big)
\eqno(A2)
$$
where $\Delta_{kk'}=\bar z^{-1}\ln k'/k-{\rm int}(\bar z^{-1}\ln k'/k)$ is the decimal part
of $\bar z^{-1}\ln k'/k$.
This equation differs from the corresponding formula for the continuous limit, Eqn. (18),
because of the exponential term in $\Delta_{\hat k\bar k}$ and $\Delta_{\hat k\bar k'}$.
This term has a fast dependence on $\hat k$, which can be treated by decomposing the
$\d\hat k$ integration as:
$
\int{\d\hat k\over\hat k}
\simeq\bar z\sum_{n_{\hat k\bar k}}\int\d\Delta_{\hat k\bar k}
$
where the integration limits in $\int\d\Delta_{\hat k\bar k}$ are approximated by $\pm\infty$
thanks to the smallness of $\Delta z$. After the Gaussian integral in $\d\Delta_{\hat k\bar k}$
is carried out, the remaining sum can be approximated back to an integral:
$\bar z\sum_{n_{\hat k\bar k}}\sim \int{\d\hat k\over\hat k}$. This means that in Eqn. (A2)
we can substitute:
$$
\exp\Big(-{\bar z^3\over\Delta z^2}\Big({\Delta_{\hat k\bar k}^2\over\ln\bar k/\hat k}+
{\Delta_{\hat k\bar k'}^2\over\ln\bar k'/\hat k}\Big)\Big)
\to\Big({\pi\Delta z^2\bar z\ln\bar k\bar k'/\hat k^2\over
\ln(\bar k/\hat k)\ln(\bar k'/\hat k)}\Big)^{1\over 2}
\exp\Big(-{\bar z^3\Delta_{\bar k\bar k'}^2\over\Delta z^2\ln\bar k\bar k'/\hat k^2}\Big).
\eqno(A3)
$$
Substituting into Eqn. (A2) and using Eqn. (17) we obtain:
$$
\langle |\Psi_{ky}|^4\rangle\simeq
{c_4\pi a_S^2\over \bar za_A^2}(kL)^{-\zeta'_4}
\left[ {a_Sa_A\over a_S^2+a_A^2}\right.
$$
$$
\left. +{2^{1\over 2}\pi\over\bar z^{3\over 2}\Delta z\hat ba_S}
\int_0^{\ln kL}\d x\Big(1+{a_A^2+a_S^2\over\hat b^2a_S^2}\e^{-2x}\Big)^{-{1\over 2}}
\Big(x+{\bar z(a_A^2+a_S^2)\over 2a_A^2a_S^2\Delta z^2}\Big)^{-{1\over 2}}
\e^{(\zeta'_4-2\zeta'_2)x}\right]
\eqno(A4)
$$
This expression differs from the continuous limit described by Eqn. (18), because of the 
factor $(x+{\bar z(a_A^2+a_S^2) \over 2a_A^2a_S^2\Delta z^2})$. If the discrete approximation
must work over the whole domain of integration in $x$, it is necessary however that 
${\bar z(a_A^2+a_S^2)\over 2a_A^2a_S^2\Delta z^2}>\ln kL$. (Incidently, this tells us that
this limit is of scarce practical interest in a probabilistic cascade, since it is not 
particularly interesting to have an error $k\Delta z$ in the cascade step, which is much 
smaller than the wavelet spectral width $k/a_S$). In the $\Delta z\to 0$ limit we get
then the following expression for the kurtosis, for $\ln RkL< (2\zeta'_2-\zeta'_4)^{-1}$:
$$
K_4(k,a_S,a_A)\simeq  {c_4\bar z\over\pi c_2^2}(kL)^{2\zeta'_2-\zeta'_4}
\left[{a_Sa_A\over a_S^2+a_A^2}+{2\pi a_Sa_AR^{2\zeta'_2-\zeta'_4}\over
\bar z^2\hat b(a_A^2+a_S)^{1\over 2}}\ln RkL\right].
\eqno(A5)
$$
Logarithmic corrections to scaling remain therefore, also in the discrete limit.

\end{document}